# Statistical Yield Modeling for IC Manufacture: Hierarchical Fault Distributions

**Yu. I. Bogdanov \*,  N. A. Bogdanova \*\*, and V. L. Dshkhunyan \***

\* OAO Angstrem, Moscow, Russia
e-mail: bogdanov@angstrem.ru
\*\* Moscow Institute of Electronic Engineering (Technical University), Moscow, Russia

*Abstract* A hierarchical approach to the construction of compound distributions for process-induced faults in IC manufacture is proposed. Within this framework, the negative binomial distribution and the compound binomial distribution are treated as level-1 models. The hierarchical approach to fault distribution offers an integrated picture of how fault density varies from region to region within a wafer, from wafer to wafer within a batch, and so on. A theory of compound-distribution hierarchies is developed by means of generating functions. With respect to applications, hierarchies of yield means and yield probability-density functions are considered and an in-process measure of yield loss is introduced. It is shown that the hierarchical approach naturally embraces the Bayesian approach.

## 1. INTRODUCTION

As early as in 1964, Murphy [1] introduced the compound Poisson distribution into microelectronics, having found that the traditional Poisson distribution is not always adequate to predict yield in integrated-circuit (IC) manufacture. The point is that process-induced faults tend to occur unevenly over the wafer, appearing as clusters. Murphy's approach was developed by Seeds [2], Okabe et al. [3], Stapper [4, 5], and other researchers.

The compound Poisson distribution differs from the traditional one in that the parameter $\lambda$ denoting the fault density is regarded as a random variable. Experience indicates that the gamma distribution is probably the most accurate model for $\lambda$ [6--10]. Also note that the compound Poisson distribution is a limiting case of the compound binomial distribution, the latter arising within Polya's urn model [11].

Statistical yield models based on compound distributions have proven to be useful in design, manufacture, and product evaluation alike. Yield enhancement aims to make the product fault-tolerant (i.e., less sensitive to process-induced faults) by adding a degree of redundancy to the IC (error-correcting codes are an example) and by optimizing its floorplan and layout [12--18]. Concerning the manufacture phase, we cite the Bayesian approach. Applied to in-process product control, the method allows one to refine yield prediction and make decisions as the batches progress along the processing line [19]. Compound-distribution models also help one to calculate yield distribution over wafers, to estimate costs, to evaluate manufacturing efficiency, to predict yield losses, etc. [20, 21].

In this study, we propose and develop a hierarchical approach to the construction of compound distributions for process-induced faults in IC manufacture.

Section 2 describes the origin and main properties of a yield model that is built around the compound Poisson distribution and has been accepted by the electronics industry [17].

Section 3 defines the compound binomial distribution and describes some of its properties. This model includes the compound Poisson distribution as a limiting case in much the same way as the traditional binomial distribution leads to the Poisson distribution. The reader is also introduced to Polya's urn model, within which the compound binomial distribution arises.

Section 4 develops a theory of compound-distribution hierarchies. Within this framework, the above-mentioned distributions belong to level 0 or 1. It is shown that main formulae can be written in compact analytical form by means of generating functions.

Section 5 deals with applied aspects. Compared with previous results, a more general formalism for mean yields is presented. Also included are equations for yield probability densities. The yield is thus treated as a hierarchical random variable.

Appendices 1 and 2 are concerned with a generalization of Polya's urn model and with the Bayesian approach to the estimation of fault density, respectively.

## 2. COMPOUND POISSON DISTRIBUTION

Primitive yield models employ the binomial distribution. Consider a chip with $n$ components. If each of them has the same probability $p$ of being faulty and faults arise independently of one another, then the probability $P(k)$ that the chip has exactly $k$ faulty components is

$$P(k) = C_n^k \cdot p^k \cdot (1-p)^{n-k}, \quad (1)$$

where

$$k = 0, 1, \ldots, n, \quad 0 < p < 1.$$

Since $n$ is very large and $p$ is very small, the binomial distribution can be approximated by a Poisson distribution to a good accuracy. Accordingly, with $\lambda$ denoting $np$ (the expected value of $k$), the probability $P(k)$ is given by

$$P(k) = \frac{\lambda^k}{k!} e^{-\lambda} \quad k = 0, 1, 2, \ldots \quad (2)$$

Assume that the chips have no redundancy, so that any faulty component will make the chip nonconforming. Then, the yield $Y$ is equal to the probability that a randomly chosen chip is fault-free:

$$Y = P(k=0) = e^{-\lambda}. \quad (3)$$

However, it was found as early as in 1964 that Eq. (3) makes badly pessimistic predictions if applied to large-area chips [1]. The underlying reason, as was revealed later, is that process-induced faults do not arise independently in different regions of the wafer but tend to cluster.

To allow for fault clustering, the compound Poisson distribution was introduced, in which the expected number of faults per chip, $\lambda$, is also a random variable. Let $P(\lambda)$ be the probability-density function (PDF) of $\lambda$. The compound Poisson distribution is defined as

$$P(k) = \int_0^\infty \frac{\lambda^k}{k!} e^{-\lambda} P(\lambda) d\lambda, \quad (4)$$

so that

$$Y = P(k=0) = \int_0^\infty e^{-\lambda} P(\lambda) d\lambda. \quad (5)$$

The density $P(k)$ might be specified in various forms. Murphy [1] proposed the triangular distribution

$$P(\lambda) = \begin{cases} \dfrac{\lambda}{\lambda_0^2} & 0 \leq \lambda \leq \lambda_0 \\ \dfrac{2\lambda_0 - \lambda}{\lambda_0^2} & \lambda_0 \leq \lambda \leq 2\lambda_0 \text{ ,} \\ 0 & \lambda > 2\lambda_0 \end{cases} \quad (6)$$

where $\lambda_0$ is the average number of faults per chip. Equations (5) and (6) imply that

$$Y = \int_0^\infty e^{-\lambda} P(\lambda) d\lambda = \left( \frac{1 - e^{-\lambda_0}}{\lambda_0} \right)^2 . \quad (7)$$

Seeds [2] assumed that

$$P(\lambda) = \frac{e^{-\lambda/\lambda_0}}{\lambda_0}, \quad (8)$$

so that

$$Y = \int_0^\infty e^{-\lambda} P(\lambda) d\lambda = \frac{1}{1 + \lambda_0} . \quad (9)$$

Okabe et al. [3] and Stapper [4, 5] specified $P(\lambda)$ as a gamma distribution:

$$P(\lambda) = \frac{b^a \lambda^{a-1} e^{-b\lambda}}{\Gamma(a)}, \quad (10)$$

where $a$ and $b$ are positive parameters and $\Gamma(a)$ is the gamma function [22, 23]. This distribution has the mean $a/b$ and the variance $a/b^2$.

Equation (4) thus becomes

$$P(k) = \int_0^\infty \frac{\lambda^k}{k!} e^{-\lambda} P(\lambda) d\lambda = \frac{\Gamma(k+a) b^a}{k! \Gamma(a)(1+b)^{k+a}} . \quad (11)$$

Distribution (11) is commonly known as the negative binomial distribution. With two parameters available for adjustment, one can fit the model to observation data in terms of variance as well as mean.

In what follows the density $P(\lambda)$ appearing in Eq. (4) will be taken in form (10). This type of compound Poisson distribution allows one to accurately represent real situations and has some theoretical advantages (see below).

The mean and variance are given by

$$\mu = \frac{a}{b}, \quad (12)$$

$$\sigma^2 = \frac{a}{b^2}(1+b) . \quad (13)$$

It is seen from Eqs. (12) and (13) that $\sigma^2 > \mu$, whereas the traditional Poisson distribution has equal mean and variance.

Solving Eqs. (12) and (13) for $a$ and $b$, we obtain

$$a = \frac{\mu^2}{\sigma^2 - \mu}, \quad (14)$$

$$b = \frac{\mu}{\sigma^2 - \mu}. \quad (15)$$

Formulas (14) and (15) are useful for fitting the distribution to a sample, with the sample mean and variance assigned to $\mu$ and $\sigma^2$, respectively.

Two limiting cases are worth noting: (i) If $a$ and $b$ tend to infinity in such a way that $a/b$ tends to a finite number $a/b \to \lambda_0 = const$, then the compound Poisson distribution approaches the traditional Poisson distribution in which $\lambda = \lambda_0$. (ii) If $a$ tends to infinity and $b$ is fixed, then the compound Poisson distribution approaches the normal distribution for which Eqs. (12) and (13) hold.

Equation (11) implies that

$$Y = P(k = 0) = \frac{1}{\left(1 + \frac{1}{b}\right)^a}. \quad (16)$$

Let us recast this in terms of the average number of faulty components per chip, $\lambda_0 = \frac{a}{b}$:

$$Y = P(k = 0) = \frac{1}{\left(1 + \frac{\lambda_0}{a}\right)^a}. \quad (17)$$

The model considered is known as the large-area clustering negative binomial model [13, 15]. The parameter $a$ is called the cluster parameter. Its typical values approximately range from 0.3 to 7. In actual fact, fault clustering disappears if $a$ exceeds 4 or 5; for such $a$, Eq. (17) can be approximated by formula (3).

The large-area clustering model is based on two assumptions. First, fault clusters are larger than chips, so that any faulty chip is totally covered by one fault cluster. Second, faults are distributed uniformly within any cluster.

In addition, there are the small-area clustering negative binomial model [24] and the medium-area clustering negative binomial model [25, 26]. The latter is regarded as including the other models [13]. It is intended for chips with areas on the order of a square inch; they may well be larger than fault clusters. Since the concept of fault cluster has yet to be clarified, the medium-area clustering model is defined in terms of blocks [25]. It is assumed that (i) correlation between faults may exist only within a block, (ii) blocks are statistically independent of each other, (iii) the total number of faults per block obeys a negative binomial distribution, and (iv) faults are distributed uniformly over each block.

### 3. COMPOUND BINOMIAL DISTRIBUTION AND POLYA'S URN MODEL

By analogy with the compound Poisson distribution, let us apply the compounding procedure to the binomial distribution. The result might be called the compound binomial distribution.

#### 3.1 Compound Binomial Distribution: Definition and Properties

Consider a population of good and faulty items in which the latter ones account for the proportion $\theta$, with the faults distributed randomly over the population. Let us take a sample of size $n$. The total number $k$ of faulty items in the sample is a random variable obeying a binomial distribution with the probability mass function (PMF)

$$P(k|\theta) = C_n^k \cdot \theta^k \cdot (1-\theta)^{n-k}, \quad (18)$$

where $k=0,1,\ldots,n$; $0<\theta<1$.

Let us regard $\theta$ as a random variable that changes from population to population. The variability of $P(\theta)$ will be modeled by a beta distribution, whose PDF is

$$P(\theta) = \frac{\Gamma(a+b)}{\Gamma(a)\cdot\Gamma(b)} \cdot \theta^{a-1} \cdot (1-\theta)^{b-1}, \quad (19)$$

where $a>0$, $b>0$ [22, 23]. The relationship between $b$ in Eq. (19) and $b$ in Eq. (10) will be dealt with in what follows.

The expected value and variance of $P(\theta)$ are given by

$$E(\theta) = \frac{a}{a+b}, \quad (20)$$

$$Var(\theta) = \frac{a\cdot b}{(a+b)^2 \cdot (a+b+1)}. \quad (21)$$

Due to Eqs. (18) and (19), the probability that a sample of size $n$ has $k$ faulty items is as follows [11]:

$$P(k) = \int_0^1 P(k|\theta) \cdot P(\theta) \cdot d\theta =$$
$$= \frac{\Gamma(n+1)\cdot\Gamma(a+b)\cdot\Gamma(k+a)\cdot\Gamma(n-k+b)}{\Gamma(a)\cdot\Gamma(b)\cdot\Gamma(n+a+b)\cdot\Gamma(k+1)\cdot\Gamma(n-k+1)} \quad (22)$$

This PMF defines the compound binomial distribution. It can also be obtained within Polya's urn model (see Subsection 3.2).

It can be shown directly or with generating functions (see Section 4) that the mean and variance of the compound binomial distribution obey the equations

$$\mu = \frac{n\cdot a}{a+b}. \quad (23)$$

$$\sigma^2 = \frac{n\cdot a\cdot b\cdot(a+b+n)}{(a+b)^2 \cdot (a+b+1)}. \quad (24)$$

Solving Eqs. (23) and (24) for $a$ and $b$, we obtain

$$a = \frac{\mu\cdot(\mu\cdot(n-\mu)-\sigma^2)}{\sigma^2\cdot n - \mu\cdot(n-\mu)}. \quad (25)$$

                                                                                                              6

$$b = \frac{(n - \mu) \cdot (\mu \cdot (n - \mu) - \sigma^2)}{\sigma^2 \cdot n - \mu \cdot (n - \mu)}. \quad (26)$$

Formulas (25) and (26) are useful for fitting PMF (22) to a sample, with the sample mean and variance assigned to $\mu$ and $\sigma^2$, respectively.

The compound binomial distribution is a more general model than the compound Poisson distribution, the binomial distribution, and the beta distribution, as evidenced by the following properties.

**Statement 1**: If $b$ and $n$ tend to infinity in such a way that $b/n$ tends to a finite number $b_0$, then the compound binomial distribution approaches the compound Poisson distribution with the parameters $a$ and $b_0$. This statement establishes a link between the $b$ parameters of the two distributions.

**Statement 2**: If $a$ and $b$ tend to infinity in such a way that $a/b$ tends to a finite number, then the compound binomial distribution approaches the binomial distribution with

$$p = \frac{(a/b)}{1 + (a/b)}.$$

**Statement 3**: If $n$ tends to infinity, whereas $a$ and $b$ are fixed, then the distribution of the random variable $k/n$ approaches the beta distribution with the parameters $a$ and $b$.

### 3.2 Polya's Urn Model

The compound binomial distribution can be obtained by a certain probabilistic algorithm in much the same way as the traditional binomial distribution is produced by Bernoulli trials. The algorithm is known as Polya's urn model. It was proposed by a Hungarian-born US mathematician in 1923 as a means of describing epidemics and similar phenomena, allowing for aftereffects [11, 27].

Polya's urn model is defined as follows. Consider an urn that initially contains $a$ black and $b$ white balls. In the microelectronics context, white and black balls might represent good and faulty IC components, respectively.

Let us repeat the following procedure continually at equal intervals:
(i) Withdraw a ball at random from the urn.
(ii) Identify the color of the ball.
(iii) Return the ball to the urn.
(iv) Add a ball of the same color to the urn.

Thus, the total number of balls in the urn steadily increases. For each cycle, the probability of retrieving a ball of given color depends on the current numbers of black and white balls in the urn. In the first cycle the probability $p$ of taking a black ball is

$$p = \frac{a}{a + b}$$

and that for a white ball is

$$q = 1 - p = \frac{b}{a + b}.$$

If black balls have appeared $k$ times in $n$ cycles (so that white ones has appeared $n - k$ times), then the probability of chosing a black ball in the $(n+1)$th cycle is

$$p_{n+1} = \frac{a + k}{a + b + n}. \quad (27)$$

Let us recast this in a more convenient form:
$$p_{n+1} = \frac{p + k \cdot g}{1 + n \cdot g}, \quad (28)$$
where
$$p = \frac{a}{a+b} \quad g = \frac{1}{a+b}.$$
It will be shown in what follows that the probability of taking a black ball in the first cycle, $p$, is equal to the average proportion of black balls.

The parameter $g$ will be called the clustering factor. Aftereffects disappear in the limit $g \to 0$. In other words, the probability of taking a black ball in an $(n+1)$th cycle is independent of the history, $p_{n+1} = p$, so that Polya's urn model will be equivalent to Bernoulli trials and $k$ will obey the binomial distribution.

Strictly speaking, Polya's urn model changes into Bernoulli trials if
$$a \to \infty, \quad b \to \infty, \quad \frac{a}{a+b} \to p = const.$$
Indeed, the changeover also occurs if one fails to add balls to the urn.

If $g > 0$ is a finite number, the probability of taking a black ball increases as $k$. Moreover, the following statements hold: (i) If $k = p \cdot n$, then $p_{n+1} = p$. (ii) If $k > p \cdot n$, then $p_{n+1} > p$. (iii) If $k < p \cdot n$, then $p_{n+1} < p$. We see that faults are more likely to be formed in regions where the total number of previously induced faults exceeds the expected value. Fault clustering thus arises. In fact, good components may also form clusters in regions where fault count is less than the expected value; this phenomenon will be dealt with in what follows.

Let us generalize the binomial distribution to the case of a finite $g > 0$. To this end, we will derive an expression for the probability of taking a black ball exactly $k$ times in $n$ cycles in the context of Polya's urn model. The probability that black balls appear in and only in the first $k$ cycles is given by
$$\frac{a(a+1)(a+2)\cdots(a+k-1) \cdot b(b+1)(b+2)\cdots(b+n-k-1)}{(a+b)(a+b+1)(a+b+2)\cdots(a+b+n)}. \quad (29)$$
Furthermore, any other sequence of $n$ cycles in which black balls appear exactly $k$ times has the same probability. The total possible number of such sequences is equal to $C_n^k$, the number of ways of selecting $k$ objects out of $n$.

Thus, the probability of taking a black ball $k$ times in $n$ cycles is given by
$$P(k|n,a,b) = \frac{\Gamma(n+1) \cdot \Gamma(a+b) \cdot \Gamma(k+a) \cdot \Gamma(n-k+b)}{\Gamma(a) \cdot \Gamma(b) \cdot \Gamma(n+a+b) \cdot \Gamma(k+1) \cdot \Gamma(n-k+1)}. \quad (30)$$
In writing this formula, we capitalized on the following property of the gamma function:
$$\frac{\Gamma(a+k)}{\Gamma(a)} = a(a+1)\cdots(a+k-1), \quad (31)$$
so that
$$\Gamma(k+1) = k!.$$

Changing from $a$ and $b$ to $p = \dfrac{a}{a+b}$ and $g = \dfrac{1}{a+b}$, we obtain

$$P(k|n,g,p) = \frac{\Gamma(n+1)\cdot\Gamma\left(\dfrac{1}{g}\right)\cdot\Gamma\left(k+\dfrac{p}{g}\right)\cdot\Gamma\left(n-k+\dfrac{1-p}{g}\right)}{\Gamma\left(\dfrac{p}{g}\right)\cdot\Gamma\left(\dfrac{1-p}{g}\right)\cdot\Gamma\left(n+\dfrac{1}{g}\right)\cdot\Gamma(k+1)\cdot\Gamma(n-k+1)} \quad (32)$$

and recast Eqs. (23) and (24) in the form

$$\mu = np, \quad (33)$$

$$\sigma^2 = npq\,\frac{(a+b+n)}{(a+b+1)} = npq\,\frac{1+gn}{1+g}. \quad (34)$$

Notice that the mean is the same as that for the binomial distribution. However, the variance of compound binomial distribution is always larger than the variance of ordinary (non-compound) binomial distribution, as with the compound Poisson distribution. The formula for the variance changes into its binomial-distribution counterpart only in the limit

$n \ll (a+b), \quad (a+b) \gg 1$.

If $n \gg (a+b)$, the variance varies as $n^2$ (recall that $\sigma^2 = npq$ for the binomial distribution).

Polya's urn model allows us to put a microelectronics interpretation on the parameters $a$ and $b$ of a gamma or a beta distribution. The parameter $b$ (the initial number of white balls) can be seen as the size of the fault-free part. With a compound binomial distribution or a compound Poisson distribution, the size is measured by the total number of good chip components or good chips, respectively. In the latter case the size can be approximated by $b/n$ (see Statement 1 of Subsection 3.1). Also note that $b$ might be replaced with $b/S$, where $S$ is the chip area. The parameter $a$ takes the same value in the two cases.

## 4. COMPOUND-DISTRIBUTION HIERARCHIES

Using generating functions, one can visually construct a hierarchy of compound distributions that represents the multilevel nature of fault clustering. We start with the generating function of a binomial distribution, which should be regarded as a compound distribution of level 0. Then we explicitly construct levels 1 and 2. Finally, we generalize the results to an arbitrary level.

### 4.1 Generating Function of a Compound Binomial Distribution: The Construction of Levels 1 and 2

Consider a binomial distribution of a random variable $k$ with the PMF

$$P(k|\theta, n) = C_n^k \cdot \theta^k \cdot (1-\theta)^{n-k}, \quad (35)$$

where

$k = 0, 1, \ldots, n, \quad 0 < \theta < 1$.

It can be expressed in terms of conditional probabilities. In the general case the generating function is defined as

$$G(z|\theta, n) = \sum_{k=0}^{n} z^k \cdot P(k|\theta, n). \quad (36)$$

Due to the Binomial Theorem, the generating function of the binomial distribution takes the form

$$G(z|\theta,n) = \sum_{k=0}^{n} z^k C_n^k \cdot \theta^k \cdot (1-\theta)^{n-k} = (1-\theta(1-z))^n \quad (37)$$

Now, let us construct level 1 of a compound-distribution hierarchy representing fault clustering. Let $\theta$ be a random variable with a beta distribution. Accordingly, we deal with three parameters: $n$, $a$, and $b$. The generating function is given by

$$G(z|n,a,b) = \frac{\Gamma(a+b)}{\Gamma(a)\Gamma(b)} \int_0^1 \theta^{a-1} \cdot (1-\theta)^{b-1} \cdot (1-\theta(1-z))^n \, d\theta \quad (38)$$

The integral representation of the hypergeometric function [28, 29] enables us to put Eq. (38) into a neat form:

$$G(z|n,a,b) = F(-n, a, a+b, (1-z)) \quad (39)$$

This is the generating function of the compound binomial distribution. The hypergeometric series is basically an $n$ th-degree polynomial in $1-z$. It is convenient to switch from $1-z$ to $z$ so that the polynomial coefficients be directly linked to probabilities. We thus obtain

$$G(z|n,a,b) = \frac{\Gamma(a+b)\Gamma(b+n)}{\Gamma(a+b+n)\Gamma(b)} F(-n, a, -n-b+1, z) \quad (40)$$

Changing from $a$ and $b$ to

$$g = \frac{1}{a+b}$$

and

$$p = \frac{a}{a+b},$$

we transform Eqs. (39) and (40) into

$$G(z|n,g,p) = F\left(-n, \frac{p}{g}, \frac{1}{g}, (1-z)\right) \quad (41)$$

and

$$G(z|n,g,p) = \frac{\Gamma\left(\frac{1}{g}\right)\Gamma\left(\frac{1-p}{g}+n\right)}{\Gamma\left(\frac{1}{g}+n\right)\Gamma\left(\frac{1-p}{g}\right)} F\left(-n, \frac{p}{g}, -n-\frac{1-p}{g}+1, z\right). \quad (42)$$

Now, let as regard the parameter $p$ as a random variable having the beta distribution with parameters $a$ and $b$ (not to be confused with the level-1 parameters of the same names). The generating function is

$$G(z|n,g,a,b) = \frac{\Gamma(a+b)}{\Gamma(a)\Gamma(b)} \int_0^1 G(z|n,g,p) \cdot p^{a-1}(1-p)^{b-1} \, dp \quad (43)$$

The PMF is given by

$$P(k|n,g,a,b) = \frac{\Gamma(a+b)}{\Gamma(a)\Gamma(b)} \int_0^1 P(k|n,g,p) \cdot p^{a-1}(1-p)^{b-1} \, dp \quad (44)$$

**4.2 Generalization: Generating-Function Hierarchy for a Compound Binomial Distribution**

Let us summarize and generalize the procedure described in the preceding subsection. At level 0, we have a binomial distribution. It is specified by a single parameter, $p_0$, which might be viewed as the proportion of faulty components. In passing to level 1, we declare the parameter $p_0$ a random variable with a beta distribution and eliminate $p_0$ by the compounding procedure defined in Subsection 3.1. As a result, we deal with parameters $a_1$ and $b_1$ instead of $p_0$. Finally, we introduce

$$g_1 = \frac{1}{a_1 + b_1}$$

and

$$p_1 = \frac{a_1}{a_1 + b_1}.$$

In general, level $r+1$ is constructed from level $r$ by treating the parameter $p_r$ as a random variable obeying the beta distribution with parameters $a_{r+1}$ and $b_{r+1}$; when passing to level $r+1$, the variable $p_r$ is replaced with the parameters $p_{r+1}$ and $g_{r+1}$ by the compounding procedure. For each $r$,

$$g_r = \frac{1}{a_r + b_r}$$

and

$$p_r = \frac{a_r}{a_r + b_r}.$$

If one progresses from level $r$ to level $r+1$, the generating function and PMF are transformed as follows:

$$G_{r+1}(z|n, g_1, g_2, ..., g_r, a_{r+1}, b_{r+1}) =$$
$$\frac{\Gamma(a_{r+1} + b_{r+1})}{\Gamma(a_{r+1})\Gamma(b_{r+1})} \int_0^1 G_r(z|n, g_1, g_2, ..., g_r, p_r) \cdot p_r^{a_{r+1}-1}(1 - p_r)^{b_{r+1}-1} dp_r, \quad (45)$$

$$P_{r+1}(k|n, g_1, g_2, ..., g_r, a_{r+1}, b_{r+1}) =$$
$$\frac{\Gamma(a_{r+1} + b_{r+1})}{\Gamma(a_{r+1})\Gamma(b_{r+1})} \int_0^1 P_r(k|n, g_1, g_2, ..., g_r, p_r) \cdot p_r^{a_{r+1}-1}(1 - p_r)^{b_{r+1}-1} dp_r. \quad (46)$$

**4.3 Generating-Function Hierarchy for a Compound Poisson Distribution**

For each level $r$ of the hierarchy, a compound Poisson distribution arises if $p_r \to 0$, $g_r \to 0$, and $n \to \infty$ in such a way that $np_r \to \lambda_r = const$ and $ng_r \to \lambda_r / a_r = const$, where $\lambda_r$ and $\lambda_r / a_r$ are finite numbers. In the microelectronics context, $n \geq 10^6$ and $p_r \leq 10^{-6}$, which refer to IC complexity and the probability of an individual IC component being faulty, respectively. Thus, the above conditions are fulfilled.

At level 0, the binomial generating function changes into the Poisson generating function:

$$G_0(z|p_0,n) = (1 - p_0(1-z))^n = \left(1 - \frac{np_0}{n}(1-z)\right)^n \to \exp(-np_0(1-z)). \quad (47)$$

For levels 0--2, the natural logarithms of the generating functions are given by

$$\ln G_0(z|n, p_0) = -np_0(1-z), \quad (48)$$

$$\ln G_1(z|n, g_1, p_1) = -\frac{p_1 n}{g_1 n}\ln(1 + g_1 n(1-z)), \quad (49)$$

$$\ln G_2(z|n, g_1, g_2, p_2) = -\frac{p_2 n}{g_2 n}\ln\left(1 + \frac{g_2 n}{g_1 n}\ln(1 + g_1 n(1-z))\right). \quad (50)$$

The factors $n$ are retained in fractions in order to show that indefinitely small and large quantities appear as products only.

In general,

$$\ln G_r = -p_r n \cdot L_r. \quad (51)$$

where

$$L_{r+1} = \frac{\ln(1 + g_{r+1} n L_r)}{g_{r+1} n}, \quad L_0 = (1-z). \quad (52)$$

The clustering factor $g_r$ varies with $r$. Concerning the probability $p_r$, it is associated with the highest level and is evaluated by averaging over all the levels. Accordingly, one could simply write $p$ instead of $p_r$.

### 4.4 PMF and Moments

A generating function contains the complete information on the random variable. Consider a discrete random variable with the generating function $G(z)$. The probability that the random variable takes a value $k$ can be expressed in terms of the $k$ th derivative of $G(z)$ at $z=0$, and an $m$ th factorial moment is equal to the $m$ th derivative of $G(z)$ at $z=1$:

$$P(k) = \frac{1}{k!}\frac{\partial^k G(z)}{\partial z^k}\bigg|_{z=0}, \quad (53)$$

$$E[k(k-1)...(k-m+1)] = \frac{\partial^m G(z)}{\partial z^m}\bigg|_{z=1}. \quad (54)$$

In particular, the expected value and variance of a discrete random variable $k$ are given by

$$E[k] = G'(1), \quad (55)$$

$$Var[k] = G''(1) + G'(1) - G'^2(1). \quad (56)$$

The above formulae imply that the mean and variance for the compound distribution of level $r$ can be expressed in simple form:

$$\mu = np, \quad (57)$$

$$\sigma^2 = np(1 + g_1 n + g_2 n + .... + g_r n). \quad (58)$$

In Eqs. (57) and (58), we omitted the subscript $p_r$ on $p$.

It follows that the clustering factors associated with the levels of the hierarchy additively contribute to the total variance. The expansion of $\sigma^2$ in terms of hierarchy levels might be useful for estimating $g_i$ from empirical data. Concerning IC manufacture, formulae (57) and (58) make it possible to develop the hierarchical analysis of variance for process-induced faults by analogy with that for product variables [30].

Even if an analytical formula is available for $G(z)$, the calculation of $P(k)$ from Eq. (53) is likely to be a time-consuming procedure, so it would be wise to perform numerical differentiation with a computer; an example is given in Table 1.

Table 1. Examples of distributions for different levels of modeling, with $n = 10^6$ and $p = 5 \cdot 10^{-7}$

| $k$ | Level 0 (Poisson model, $g_1 = g_2 = 0$) | Level 1 ($g_1 = 10^{-6}, g_2 = 0$) | Level 2 ($g_1 = g_2 = 10^{-6}$) |
|---|---|---|---|
| 0 | 0.6065 | 0.7071 | 0.7685 |
| 1 | 0.3033 | 0.1768 | 0.1135 |
| 2 | 0.0758 | 0.0663 | 0.0535 |
| 3 | 0.0126 | 0.0276 | 0.0282 |
| 4 | 0.0016 | 0.0121 | 0.0155 |
| 5 | 0.0002 | 0.0054 | 0.0088 |
| 6 | 0.0000 | 0.0025 | 0.0050 |
| 7 | 0.0000 | 0.0012 | 0.0029 |
| 8 | 0.0000 | 0.0005 | 0.0017 |
| 9 | 0.0000 | 0.0003 | 0.0010 |
| 10 | 0.0000 | 0.0001 | 0.0006 |
| 11 | 0.0000 | 0.0001 | 0.0003 |
| 12 | 0.0000 | 0.0000 | 0.0002 |
| 13 | 0.0000 | 0.0000 | 0.0001 |
| 14 | 0.0000 | 0.0000 | 0.0001 |
| 15 | 0.0000 | 0.0000 | 0.0000 |

Table 1 indicates that hierarchical clustering models predict a higher yield than simple Poisson models. At the same time, the former distributions show longer tails. Numerical calculations demonstrate that compound Poisson distributions lead to estimates very close to those obtained by more laborious manipulations on the basis of compound binomial distributions (see Eqs. (22) and (44)).

## 5. YIELD ANALYSIS
### 5.1 Yield Hierarchy

In the absence of redundancy the yield is measured by the probability of choosing a fault-free chip ($k = 0$); therefore, it is simply equal to the value of the generating function at $z = 0$. Thus, for chips of complexity $n$, the generating-function hierarchy can easily be associated with a yield hierarchy. With $\ln Y_0 = -np_0$, we have

$$Y_0 = \exp(-np_0). \quad (59)$$

For levels 1 and 2,

$$\ln Y_1 = -p_1 n \frac{\ln(1 + g_1 n)}{g_1 n}, \quad (60)$$

$$\ln Y_2 = -p_2 n \frac{\ln\left(1 + g_2 n \frac{\ln(1+g_1 n)}{g_1 n}\right)}{g_2 n}. \quad (61)$$

In general,

$$\ln Y_r = -p_r n \cdot L_r, \quad (62)$$

where

$$L_{r+1} = \frac{\ln(1 + g_{r+1} n L_r)}{g_{r+1} n}, \quad L_0 = 1. \quad (63)$$

Comparing these with Eqs. (51) and (52), we see that

$$L_r \equiv L_r(z=0).$$

To avoid confusion, we did not introduce new letters.

Table 2 gives examples of yield hierarchies for different IC complexities. The data reflect the fact that the Poisson model without clustering badly underestimates the yield of high-complexity ICs. Also notice that clustering is stronger in a level-2 model, provided that $g_1 = g_2$.

| Table 2. Yield as a function of IC complexity for different levels of modeling ($p = 10^{-7}$) | | | | | |
|---|---|---|---|---|---|
| IC complexity | 256 K | 1 M | 4 M | 16 M | 64 M |
| Level 0 (Poisson model, $g_1 = g_2 = 0$) | 0.9741 | 0.9005 | 0.6574 | 0.1868 | 0.0012 |
| Level 1 ($g_1 = 5 \cdot 10^{-7}, g_2 = 0$) | 0.9757 | 0.9192 | 0.7976 | 0.6390 | 0.4924 |
| Level 2 ($g_1 = g_2 = 5 \cdot 10^{-7}$) | 0.9770 | 0.9321 | 0.8596 | 0.7905 | 0.7388 |

### 5.2 Yield as a Random Variable: Its Distribution

Let us consider a block (i.e., a limited area) on a silicon wafer [25]. With $p_0$ denoting the fault density for the block, the yield is given by the Poisson formula

$$Y_0 = \exp(-np_0).$$

Further, $p_0$ itself varies randomly from block to block according to a beta distribution. (In the situation considered, the beta distribution can be approximated by a gamma distribution to a very high accuracy.)

Let us change from the random variable $p_0$ to the random variable $Y_0 = \exp(-np_0)$. The PDF of the latter is expressed as follows [21]:

$$P(Y_0) = \frac{(b_1/n)^{a_1}}{\Gamma(a_1)} (-\ln Y_0)^{a_1 - 1} \cdot Y_0^{(b_1/n - 1)}. \quad (64)$$

In general,

$$Y_r = \exp(-p_r n \cdot L_r), \quad (65)$$

so that

$$P(Y_r) = \frac{1}{\Gamma(a_{r+1})} \left(\frac{b_{r+1}}{nL_r}\right)^{a_{r+1}} (-\ln Y_r)^{a_{r+1}-1} Y_r^{\left(\frac{b_{r+1}}{nL_r}-1\right)}, \quad (66)$$

$$P(Y_r) = \frac{1}{\Gamma\left(\frac{p_{r+1}}{g_{r+1}}\right)} \left(\frac{1}{g_{r+1} nL_r}\right)^{\frac{p_{r+1}}{g_{r+1}}} (-\ln Y_r)^{\frac{p_{r+1}}{g_{r+1}}-1} Y_r^{\left(\frac{1}{g_{r+1} nL_r}-1\right)}. \quad (67)$$

Here, $Y_r$ is the yield (average value) for level $r$. In passing to level $r+1$, one treats $Y_r$ as a random variable with the PDF $P(Y_r)$. This approach enables one to naturally describe the random variation of yield from block to block within a wafer, from wafer to wafer within a batch, and so on. Figure 1 shows yield PDFs for different IC complexities. The parameter $g$ refers to fault clustering within a wafer, whereas $a$ and $b$ serve to allow for the nonuniformity of fault distribution over the wafers.

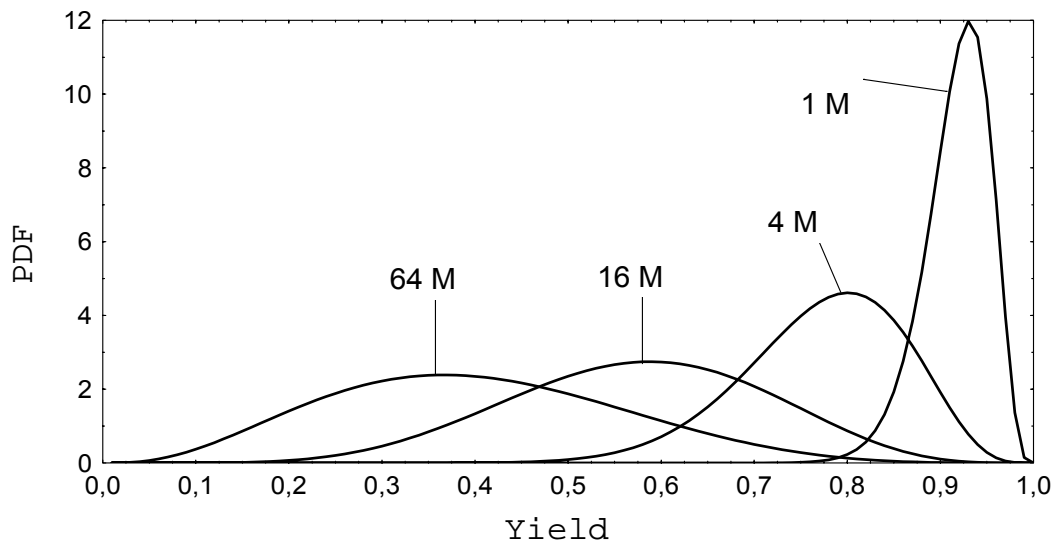

Fig. 1. Yield PDFs for different IC complexities
(g=3*10$^{-7}$, a=5, fault density D=a/(a+b)=10$^{-7}$)

**5.3 In-process Measure of Yield Loss**
As an in-process measure of yield loss, one might take the cumulative distribution function (CDF) of yield, $F(Y_r)$, using Eq. (66) [20, 21]. The CDF is expressed by means of incomplete gamma function as

$$F(Y_r) = 1 - \gamma\left(-\frac{b_{r+1}}{nL_r} \ln Y_r; a_{r+1}\right), \quad (68)$$

where:

$$\gamma(x;a) = \frac{1}{\Gamma(a)} \int_0^x t^{a-1} \cdot e^{-t} dt \quad (69)$$

Figure 2 illustrates the approach in the case of intolerable yield loss. The solid and the broken curve depict a specified and an actual CDF, respectively. The yield and yield loss are measured by the areas of the corresponding parts of the square: the part to the right of the specified curve refers to the specified yield loss, the part to the left of the actual curve represents the actual yield, and the part between the actual and the specified curve indicates the excess yield loss. The process should be controlled in such a way that the actual curve mostly go below the specified one. It is convenient to measure the statistical significance of excess yield loss by the Kolmogorov statistic $D^+$ [22, 23]. Moreover, the significance of the $D^-$ statistic indicates that the process capability may have been underestimated.

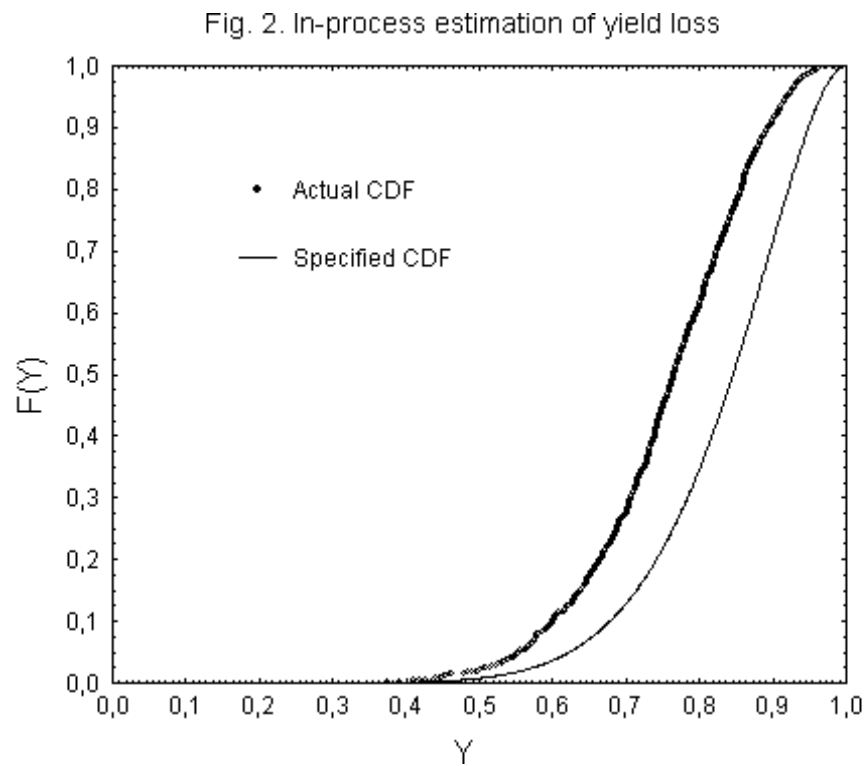

Fig. 2. In-process estimation of yield loss

## 6. CONCLUSIONS

We have developed a general approach to the construction of compound distributions for process-induced faults in IC manufacture. It quantifies the performance of a process in hierarchical form and helps one build general statistical models for fault control and yield management. With the aid of generating functions, main results can be written in compact analytical form. The strategy covers the negative binomial distribution and the compound binomial distribution, regarding them as level-1 models. (The latter distribution arises within Polya's urn model.)

The hierarchical approach to fault distribution offers an integrated picture of how fault density varies from area to area within a wafer, from wafer to wafer within a batch, and so on. The average fault density and the clustering factor can be estimated from an expansion of the variance in terms of hierarchy levels.

Yield hierarchies provide a new, general formalism for yield estimation and prediction in IC manufacture, treating the yield as a random variable. Particular cases are the Poisson distribution (level-0 model) and the negative binomial distribution (level-1 model).

The in-process measure of yield loss proposed in this study might be useful for setting standards and implementing statistical process control in real time.

Finally, two points are worth making:

(i) Polya's urn model can be generalized to the case of noninteger parameters and an arbitrary number of hierarchy levels (see Appendix 1).

(ii) The hierarchical approach to fault distribution naturally embraces the Bayesian approach (see Appendix 2).

**APPENDIX 1: A GENERALIZATION OF POLYA'S URN MODEL**

In this appendix we put forward a generalization of Polya's urn model that (i) permits $a$ and $b$ to take noninteger values and (ii) naturally leads to the distribution hierarchy. This generalization might be called the sandbox model.

Consider a sandbox that initially contains $a$ cups of dark-colored sand and $b$ cups of light-colored sand. The two types of sand are thoroughly mixed, and their grains are of the same size.
Let us perform this Procedure:
(i) Take a grain at random from the sandbox.
(ii) Identify the color of the grain.
(iii) Add a cup of sand of the same color to the sandbox.
(iv) Stir the sand in the sandbox.

After each cycle, the proportion $p$ of dark sand in the sandbox is a random variable. Its distribution will approach the beta distribution with parameters $a$ and $b$ if the Procedure is repeated indefinitely. Effectively, we deal with a population of sandboxes whose states are quantified by $p$. During a sufficiently large number of cycles, the sandboxes will evolve from the same initial state, characterized by $a$ and $b$, to different final states obeying the beta distribution with the parameters $a$ and $b$ (see Subsection 3.2). A uniform population of sandboxes has thus become randomly nonuniform in $p$ according to the beta distribution. On the other hand, if we take a sample of $n$ from each sandbox, the samples will make up a population with the same distribution as that produced by Polya's urn model.

Let us now construct a distribution hierarchy with levels 0, 1, 2, …, $l$, starting from level $l$. For this level, we have a population of sandboxes each of which initially contains $a_l$ and $b_l$ cups of dark and light sand, respectively. To this population, we continually apply the Procedure until $p$ attains the beta distribution with the parameters $a_l$ and $b_l$.

Level $l-1$ is formed as follows. From each sandbox, let us take a number of samples of the same size. Any two samples have the same $p$ if they correspond to the same sandbox. However, $p$ varies randomly from sample to sample according to the beta distribution if no two samples considered are taken from the same sandbox. Let us regard each sample as a sandbox of level $l-1$, which contains $a_{l-1}$ cups of dark sand and $b_{l-1}$ cups of light sand. To each sandbox population uniform in $p$, we continually apply the Procedure until $p$ attains the beta distribution with the parameters $a_{l-1}$ and $b_{l-1}$.

The lower levels are created in the same way. The point is that, at each level higher than level 0, we transform a sandbox population uniform in $p$ into a randomly nonuniform one, which is characterized by a beta distribution. At level 0, we simply take samples of the same size from the sandboxes of level 1.

Let us put a microelectronic interpretation on the above hierarchy. Level 0 corresponds to chips within a block, i.e., a uniform region on the wafer. For any block, the yield is modeled by the binomial distribution with an appropriate $p$. Level 1 corresponds to blocks within a wafer. The beta distribution of this level governs the random fault variability from block to block. Level 2 represents the fault variability from wafer to wafer within a batch. Level 3 describes the variability from batch to batch within a group of batches, and so on.

**APPENDIX 2: INCORPORATING THE BAYESIAN APPROACH**

Below we demonstrate that the formalism developed in this work can naturally incorporate the Bayesian approach as an efficient method for improving statistical estimates [19]. Consider a wafer regarded as consisting of $n$-component blocks. The total number $k$ of faulty components in a block is a random variable whose distribution changes from block to block. Let $E(k)$ and $Var(k)$ be respective estimates of the expected value and variance for $k$. If there is considerable clustering, then $Var(k) > E(k)$. If faults obey a beta distribution, its parameters $a$ and $b$ can be estimated from Eqs. (25) and (26). If the difference between the mean and variance is not statistically significant, we can infer that $a$ and $b$ are indefinitely large and the compound (binomial or Poisson) distribution turns into a pure one. This implies that the wafer is of uniform fault density and can be regarded as a single block, with the fault density estimated by taking the ratio of the total number of faults to the total number of components; otherwise, the blocks have to be treated individually.

Bayes' Theorem enables one to improve the estimation of fault density for each block, provided that the data on the fault count for the block are used in conjunction with the parameters $a$ and $b$ of the interblock fault distribution. Specifically, the fault density $\theta$ of a block is regarded as a random variable whose PDF is calculated from

$$P(\theta|k) = \frac{P(k|\theta)P(\theta)}{\int P(k|\theta)P(\theta)\,d\theta}. \quad (70)$$

This approach leads to the following result. Let the random variability of $\theta$ from block to block within the wafer obey the beta distribution with parameters $a$ and $b$ (a prior distribution). Then the distribution of $\theta$ for an $n$-component block with $k$ faulty components is the beta distribution with the parameters $a' = a + k$ and $b' = b + n - k$ (a posterior distribution).

The above statement can easily be proven within Polya's urn model (see Eq. (27)).